\documentclass[traditabstract]{aa}

\usepackage{amssymb, amsmath}
\usepackage[below]{placeins}
\usepackage{graphicx,natbib,txfonts}
\usepackage{verbatim}
\usepackage{subfigure}
\numberwithin{figure}{section}

\renewcommand{\d}{\mathrm{d}}

\DeclareGraphicsExtensions{.ps,.eps,.pdf,.png,.jpg,.mps,.gif}
\graphicspath {{./}}

\begin{document}

\title{Reconstructing the projected gravitational potential of galaxy clusters from galaxy kinematics}

\author
 {Eleonora Sarli\inst{\ref{1}}\thanks{\email{Sarli@uni-heidelberg.de}} \and Sven Meyer\inst{\ref{1}} \and Massimo Meneghetti\inst{\ref{2}, \ref{3}} \and Sara Konrad\inst{\ref{1}} \and Charles L. Majer\inst{\ref{1}} \and Matthias Bartelmann\inst{\ref{1}}}
\institute
 {{Universit\"at Heidelberg, Zentrum f\"ur Astronomie, Institut f\"ur Theoretische Astrophysik, Philosophenweg 12, 69120 Heidelberg, Germany}\label{1}
 \and {INAF - Osservatorio Astronomico di Bologna, via Ranzani 1, 40127, Bologna, Italy}\label{2}
 \and {INFN - Sezione di Bologna, viale Berti Pichat 6/2, 40127, Bologna, Italy}\label{3}}

\date{\today}

\abstract
 {We develop a method for reconstructing the two-dimensional, projected gravitational potential of galaxy clusters from observed line-of-sight velocity dispersions of cluster galaxies. It is the third of an intended series of papers aiming at a unique reconstruction method for cluster potentials combining lensing, X-ray, Sunyaev-Zel'dovich and kinematic data. The observed galaxy velocity dispersions are deprojected using the Richardson-Lucy algorithm. The obtained radial velocity dispersions are then related to the gravitational potential by using the tested assumption of a polytropic relation between the effective galaxy pressure and the density. Once the gravitational potential is obtained in three dimensions, projection along the line-of-sight yields the two-dimensional potential. For simplicity we adopt spherical symmetry and a known profile for the anisotropy parameter of the galaxy velocity dispersions. We test the method with a numerically simulated galaxy cluster and galaxies identified 
therein. We extract a projected velocity-dispersion profile from the simulated cluster and pass it through our algorithm, showing that the deviation between the true and the reconstructed gravitational potential is $\lesssim10~\%$ within $\approx1.2\,h^{-1}\mathrm{Mpc}$ from the cluster centre.}
 

\maketitle

\section{Introduction}\label{sect:introduction}

The success of the cosmological standard model implies that massive, gravitationally bound cosmic objects such as galaxy clusters should be dominated by dark matter and characterised by its properties. Numerical simulations routinely find that the dark-matter distribution in galaxy clusters is expected to exhibit universal properties, for example its radial density profile and its degree of substructure. For our understanding of the nature of dark matter, it is important to test whether the mass distribution in real clusters confirms the expectations from simulations.

Galaxy clusters are much less affected by baryonic physics than galaxies since the baryonic cooling time exceeds their lifetime except in their cores. Thus, they represent an important probe for the nature of dark matter and play a key role in testing our current understanding of cosmic structure formation.

A growing number of increasingly wide surveys in a broad range of wavebands provide or will soon provide precise information on large galaxy-cluster samples. For instance, one of the goals of the ongoing Cluster Lensing And Supernova survey with Hubble (CLASH, \citep{postman_cluster_2012}) is the mapping of dark matter in clusters based on strong and weak gravitational lensing. The Dark Energy Survey (DES, \citep{frieman_dark_2013}), having started in September 2012, will combine several probes of Dark Energy, and ESA's Euclid mission \citep{laureijs_euclid_2011} will mainly focus on weak gravitational lensing measurements and galaxy clustering, but will also provide data on galaxy clusters and the integrated Sachs-Wolfe effect. The Kilo Degree Survey (KiDS) aims to map the matter distribution in the universe by means of weak gravitational lensing and photometric redshift measurements \citep{deJong_2013}.

Strong and weak gravitational lensing are widely used as an effective tool for the reconstruction of the projected mass or gravitational-potential distributions in galaxy clusters \citep{bartelmann_maximum-likelihood_1996, cacciato_combining_2006, merten_combining_2009, coe_clash:_2012, bradac_strong_2005, bradac_lensing_2010, meneghetti_weighing_2010, medezinski_cluster_2013}.

Lensing effects are due to light deflection only and thus (largely) insensitive to equilibrium and stability assumptions. They can be completely characterised by the scaled and projected Newtonian gravitational potential of the lensing matter distributions and thus most directly constrain the projected, two-dimensional gravitational potential. Non-parametric, adaptive methods have been developed and are now routinely being applied to recover cluster potentials. However, clusters provide a multitude of other observables through X-ray emission, the thermal Sunyaev-Zel'dovich effect and galaxy kinematics. In the series of articles which this paper belongs to, we are studying how all clusters observables could possibly be used to jointly constrain the projected gravitational potential of galaxy clusters with as little prejudice as possible. Combining all observables has the substantial advantage that all available information is bundled in single models, and that a range of linear scales covering 
approximately two orders of magnitude can faithfully be bridged.

In this paper, we extend our previous studies towards including information from galaxy kinematics. The observables here are the line-of-sight velocity dispersions of the cluster galaxies as a function of cluster-centric radius. The relation between three-dimensional galaxy velocity dispersions and the gravitational potential is governed in equilibrium by the Jeans equation. It resembles the equation of hydrostatic equilibrium, but contains an additional term taking account of a possible anisotropy in the velocity distribution. We describe the galaxy velocity dispersions here as an effective, possibly anisotropic pressure related to the matter density by an effective polytropic relation. Under this assumption, which we find well satisfied in simulations, the effective galaxy pressure is related to the gravitational potential by a Volterra integral equation of the second kind, which can be solved iteratively. We can then proceed as in the previous papers of this series: We apply the Richardson-Lucy 
deprojection to the observable line-of-sight velocity dispersions to obtain a three-dimensional effective galaxy-pressure profile. This is then converted to the three-dimensional potential, from which the two-dimensional potential is found by projection.

Our paper is organized as follows: We review in Sect.~\ref{sect:model} the basic relations underlying our analysis, in particular our treatment of the Jeans equation and the Richardson-Lucy deprojection. Numerical tests of our method based on kinematic data of a numerically simulated cluster and intended applications to observational data are described in Sect.~\ref{sect:num_tests}. We outline our conclusions in Sect.~\ref{sect:Conclusions}.

\section{Recovering the projected gravitational potential from the projected velocity dispersions}\label{sect:model}

\subsection{Basic relations}\label{subs:relations}

In order to incorporate measurements of the kinematics of cluster galaxies into reconstructions of the two-dimensional gravitational potential, we first require a relation between galaxy velocity dispersions and the three-dimensional gravitational potential. The velocity dispersions are generally defined (see \citep{binney_galactic_1987, schneider_extragalactic_2006}) as the mean squared deviations of the velocities of the cluster members from the mean velocity $\langle v_i\rangle$ of the population:
\begin{equation}
  \sigma_i^2 = \langle v_i^2\rangle-\langle v_i\rangle^2\;.
\label{eq:velocity_dispersion}
\end{equation}
Measured velocity dispersions are density-weighted projections of the three-dimensional velocity dispersions along lines-of-sight through the cluster. Throughout the paper, we orient the coordinate system such that the $z$-axis coincides with the line-of-sight. A projected velocity dispersion profile is constructed by averaging over concentric cylinderical shells drilled around the line-of-sight as a symmetry axis. This profile represents the input of our method.

The final output of our algorithm will be the projected Newtonian potential of the lens, defined by
\begin{equation}
  \psi(\vec\theta) = \int\phi(D_\mathrm{d}\vec\theta, z)\,\d z\;,
\label{eq:lensing_potential}
\end{equation}
where $\phi$ is the three-dimensional gravitational potential of the cluster at distance $D_\mathrm{d}$ from the observer. The projected potential is given as a function of the two-dimensional position coordinate $\vec\theta$ on the sky.

\begin{figure}[!ht]
 \centering
\includegraphics[width=\hsize]{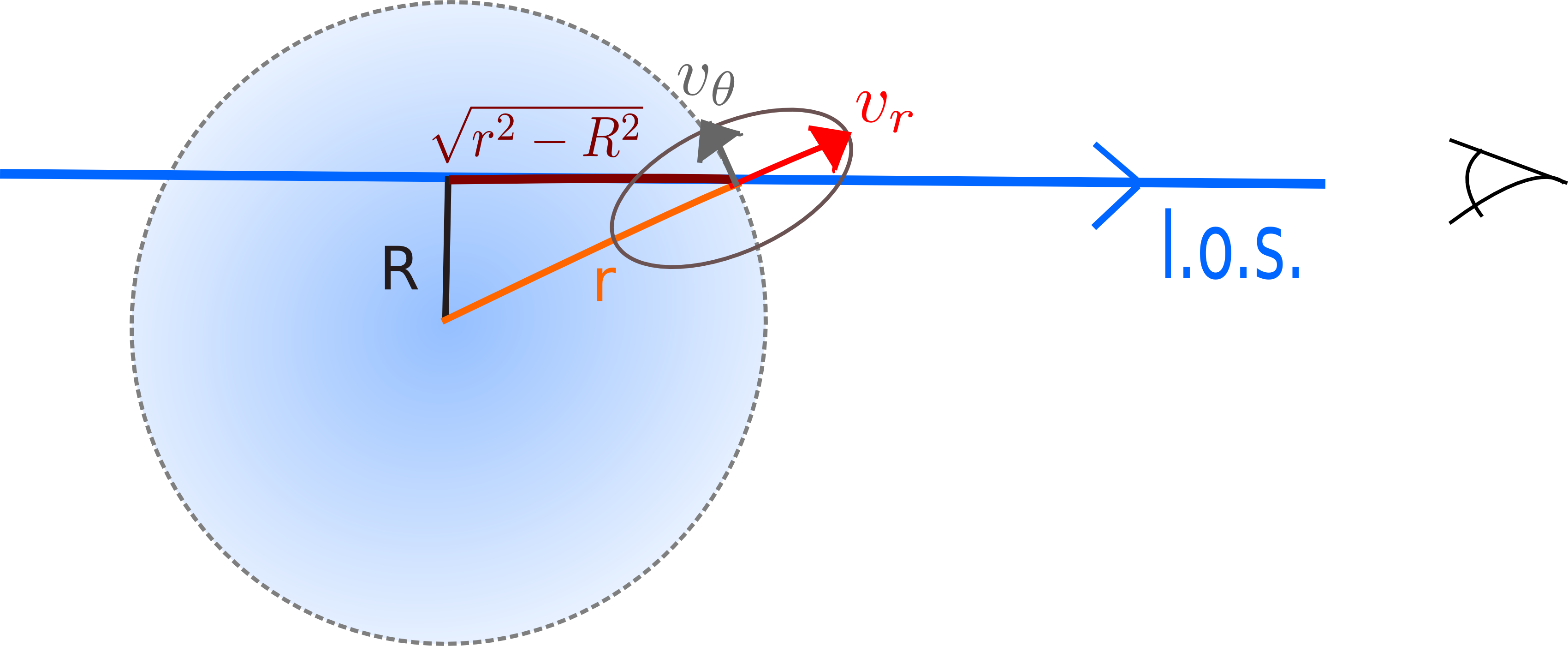} 
\caption{ Cluster geometry (see \citep{binney_galactic_1987}).}
\label{fig:cluster_geometry}
\end{figure}

The geometry of the problem is sketched in Fig.~\ref{fig:cluster_geometry} (see also \citep{binney_galactic_1987}). For simplicity during the construction of our method, we adopt a spherically-symmetric cluster model. All equations derived in the following will thus be formulated in spherical coordinates.

The possible anisotropy of the velocity distribution will be described by the conventional anisotropy parameter $\beta$ \citep{binney_galactic_1987},
\begin{equation}
 \beta = 1-\frac{\sigma_\theta^2}{\sigma_r^2}\;.
\label{eq:beta_parameter}
\end{equation}

Our algorithm consists of three main steps:

\begin{itemize}
  \item We deproject the observable, i.e.\ the velocity dispersions along the line-of-sight, into a deprojected quantity which, for reasons to be clarified later, will be called the effective galaxy pressure $P$.
  \item Since this effective pressure $P$ is related to the gravitational potential $\phi$, we can solve the Jeans equation using symmetry assumptions and a formal analogy to a polytropic gas stratification. We thus obtain a relation between the gravitational potential $\phi$ and the effective galaxy pressure. This equation is a Volterra integral equation of the second kind for $\phi$, which can be solved iteratively.
  \item Having obtained the gravitational potential, we simply project it to find the two-dimensional potential $\psi$.
\end{itemize}

We will describe our deprojection method in Subsect.~\ref{subs:deprojection}. The relevant three-dimensional quantities are the density $\rho$, the mean radial velocity dispersion $\sigma_r^2$ and the gravitational potential $\phi$. They are related via the Poisson and Jeans equations. In spherical symmetry, the Poisson and Jeans equations are
\begin{equation}
 \frac{1}{r^2}\frac{\partial}{\partial r}\left(r^2\frac{\partial\phi}{\partial r}\right)=
 4\pi G\rho(r)\;,
\label{eq:poisson}
\end{equation}
and
\begin{equation}
  \frac{1}{\rho}\frac{\partial(\rho\sigma_r^2)}{\partial r}+2\beta\frac{\sigma_r^2}{r}=
  -\frac{\partial\phi}{\partial r}\;,
\label{eq:jeans}
\end{equation}
respectively. We note here that the second term on the left-hand side of Eq.~(\ref{eq:jeans}) is the only formal difference to the hydrostatic equation for the hot intracluster gas. This difference is important because it complicates the solution of the Jeans equation considerably.

For solving the Jeans equation, we formally identify the density-weighted radial velocity dispersion $\rho\sigma_r^2$ with an effective galaxy pressure $P$ and assume that it is related to the density by a polytropic relation. This assumption can be justified by the following calculation.

A given radial density profile $\rho(r)$ implies the mass profile
\begin{equation}
  M(r) = 4\pi\int_0^r{x^2\rho(x)\d x}
\end{equation}
and, by Poisson's equation, the gravitational-potential gradient
\begin{equation}
  \frac{\partial\phi(r)}{\partial r} = \frac{GM(r)}{r^2}\;.
\end{equation}
With this expression, the Jeans equation has the solution
\begin{align}
  P(r) &= P_0\exp{\left(-\int_{r_0}^r{\frac{2\beta}{x}\d x}\right)}\nonumber\\&-
  \int_{r_0}^r{\d y\frac{GM(y)\rho(y)}{y^2}
  \exp{\left(\int_{r_0}^y{\frac{2\beta(x)}{x}\d x}\right)}}\;,
\label{eq:eff_press}
\end{align}
where the boundary conditions are set by the pressure $P_0$ at the fiducial radius $r_0$. The fiducial pressure $P_0$ can be related to a fiducial density $\rho_0$ via the virial theorem,
\begin{equation}
  \frac{P_0}{\rho_0} = \langle v^2\rangle = -\phi(r_0) \quad\Rightarrow\quad
  P_0 = \rho_0\int_\infty^{r_0}{\frac{GM(y)}{y^2}\d y}\;.
\end{equation}

\begin{figure}[!ht]
  \includegraphics[width=\hsize]{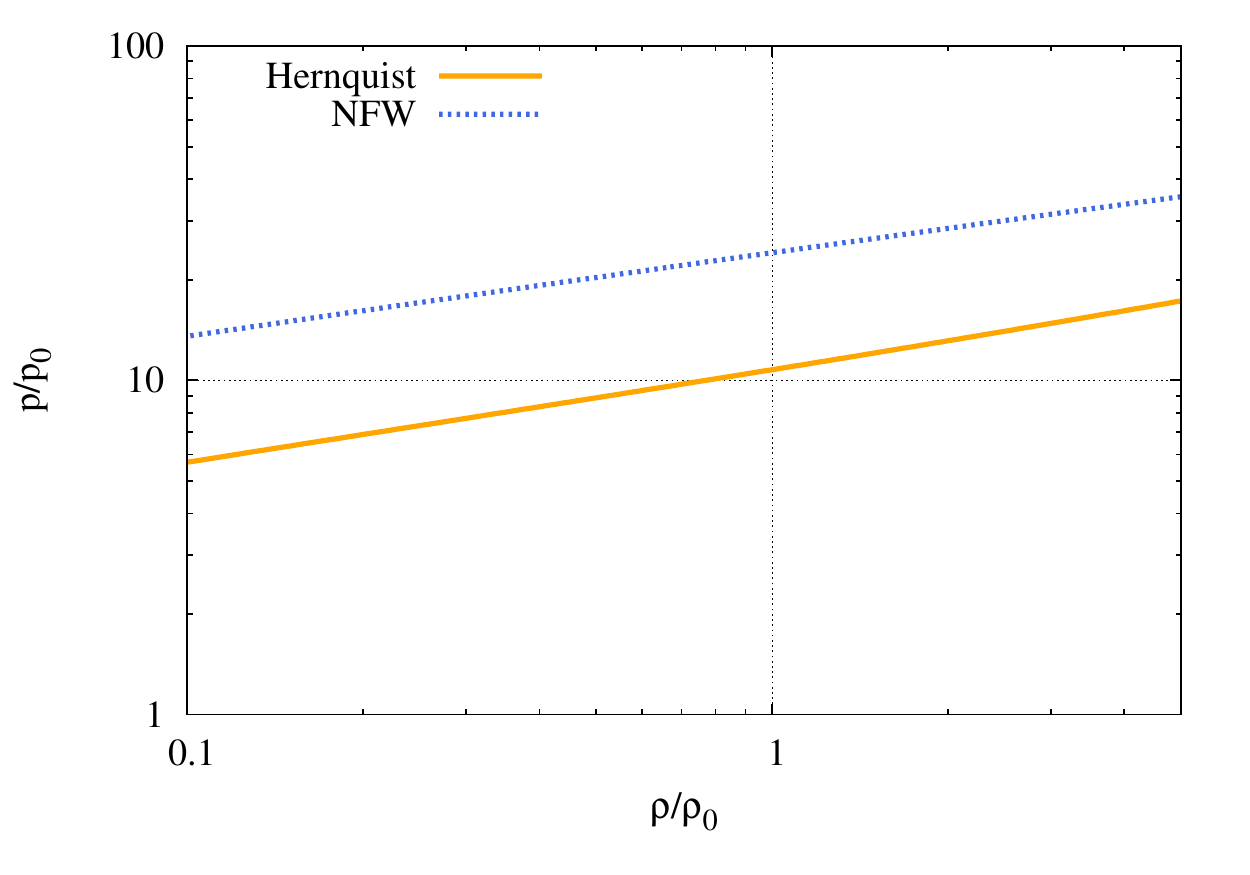} 
\caption{The relation (\ref{eq:eff_press}) between the effective pressure and the density is show here for the Hernquist and NFW density profiles, adopting the anisotropy parameter proposed by \citep{hansen_universal_2006}. The very nearly straight lines (note the logarithmic axes) demonstrate that the assumption of a polytropic relation is well justified.}
\label{fig:effective_eos}
\end{figure}

This way, the effective pressure profile of the cluster galaxies can be expressed in terms of the density and anisotropy profiles. We have evaluated Eq.~(\ref{eq:eff_press}) for two different mass density profiles, viz.\ the Hernquist \citep{hernquist_analytical_1990} and NFW \citep{navarro_structure_1996} profiles, adopting the anisotropy profile derived by \citep{hansen_universal_2006} from detailed numerical studies. As Fig.~\ref{fig:effective_eos} shows, the effective pressure-density relation is nearly polytropic, i.e.\ an approximate power law. Thus, our assumption of an effectively polytropic galaxy stratification,
\begin{equation}
  P = P_0\left(\frac{\rho}{\rho_0}\right)^\gamma\;,
\label{eq:polytropic}
\end{equation} 
seems appropriate.

With this result, we return to the Jeans equation (\ref{eq:jeans}), where we express the density by the effective pressure by means of Eq.~(\ref{eq:polytropic}). For brevity, we further abbreviate
\begin{equation}
 \epsilon = \frac{\gamma - 1}{\gamma}\;,\quad
 p = \frac{P}{P_0}\;,\quad
 \varphi = \frac{\rho_0\epsilon}{P_0}\phi\;,
\label{eq:pressure}
\end{equation}
to cast the Jeans equation (Eq.~(\ref{eq:jeans})) into the form
\begin{equation}
  \frac{\d p^\epsilon}{\d r}+\frac{2\epsilon\beta}{r}p^\epsilon =
  -\frac{\d\varphi}{\d r}\;.
\label{eq:hydrostatic2}
\end{equation}
This linear, inhomogeneous, first-order differential equation with non-constant coefficients can straightforwardly be solved, e.g.\ by variation of constants. The solution
\begin{align}
  p^\epsilon &= -\varphi(r)+
  \exp\left(-2\int_{r_0}^r\frac{\epsilon\beta}{x}\d x\right)\nonumber\\&+
  2\int_{r_0}^r\d y\frac{\epsilon\beta}{y}\varphi(y)
  \exp\left(2\int_r^y\frac{\epsilon\beta}{x}\d x\right)
\label{eq:Volterra} 
\end{align}
is the unique relation between the effective pressure and the gravitational potential we were aiming at.

Equation (\ref{eq:Volterra}) is a Volterra integral equation of second kind \citep{abramowitz_handbook_1972} that can be solved iteratively. In this way, we find a relation between the effective pressure $p$ and the gravitational potential \textbf{$\varphi$}. Projection along the line-of-sight leads to the two-dimensional potential $\psi$.

\subsection{Deprojection}
\label{subs:deprojection}

Observations yield the line-of-sight projection of the galaxy velocity dispersion only. Three-dimensional velocity dispersions can be obtained by deprojection, which can be achieved e.g.\ by Richardson-Lucy deconvolution \citep{richardson_bayesian-based_1972, lucy_iterative_1974}. As detailed in \citep{konrad_X_rays_2013}, a line-of-sight-projection $g(s)$ is assumed to be related to a three-dimensional quantity $f(r)$ by
\begin{equation}
  g(s) = \int\d z\,f\left(\sqrt{s^2+z^2}\right) = \int\d r\,f(r)K(s|r)
\label{eq:g}
\end{equation}
with a projection kernel $K(s|r)$. In spherical symmetry, with the anisotropy parameter $\beta(r)$, the projection kernel for the velocity dispersion is
\begin{equation}
  K(s|r) = \frac{2}{\pi}\frac{r}{\sqrt{r^{2}-s^2}}\Theta(r^{2}-s^2)
  \left(1-\beta(r)\frac{s^2}{r^2}\right)\;.
\label{eq:kernel}
\end{equation}
For isotropic velocity dispersions or an isotropic gas pressure, the final factor in parentheses in Eq.~(\ref{eq:kernel}) is unity. The kernel in Eq.~(\ref{eq:kernel}) can be easily derived from Fig.~\ref{fig:cluster_geometry} as shown by \citep{binney_galactic_1987}.

Provided that $g(s)$, $f(r)$ and $K(s|r)$ are normalised\footnote{The kernel $K(s|r)$ is normalised integrating over $s$.}, the following iterative scheme follows from Bayes' theorem (see \citep{lucy_iterative_1974, konrad_X_rays_2013}):
\begin{equation}
  \tilde f_{i+1}(r) = \tilde f_i(r)\int\d s\,\frac{g(s)}{\tilde g_i(s)}K(s|r)\;,
\label{lucy1}
\end{equation}
where
\begin{equation}
  \tilde g_i(s) = \int\d r\,K(s|r)\tilde f_i(r)\;.
\label{lucy2}
\end{equation}

Thus, starting from a suitably guessed, normalised function $\tilde f_0(r)$, the scheme given by Eqs.~(\ref{lucy1}) and (\ref{lucy2}) allows to recover the three-dimensional function $f(r)$ from its two-dimensional projection $g(s)$, assuming the symmetry incorporated into the projection kernel $K(s|r)$. Experience shows that even for guess functions $\tilde f_0(r)$ wildly different from the true $f(r)$, the method converges surprisingly quickly.

For applying this approach to realistic observational data containing small scale fluctuations due to background or instrumental noise, a regularisation term should be introduced as discussed in \citep{lucy_optimum_1994}. It was shown there that the deprojection algorithm can be interpreted as the result of a maximum-likelihood problem obtained by variation,
\begin{equation}
\label{var1}
\begin{split}
  f_{i+1}(r) = f_i(r) + f_i(r)\left[
    \frac{\delta H[f_i]}{\delta f_i(r)}-
    \int\d r\,f_i(r)\frac{\delta H[f_i]}{\delta f_i(r)}
  \right]\;,
\end{split}
\end{equation}
where the functional derivatives of the functional $H[f]$
\begin{equation}
  H[f] = \int\d s\,g(s)\ln\tilde g(s)
\end{equation}
occur.

Regularisation can be introduced by adding a functional $S[f]$,
\begin{equation}
  H[f] \to Q[f] = H[f] + \alpha S[f]\;,
\label{eq:regularised_functional}
\end{equation}
which can be of entropic form
\begin{equation}
  S[f] = -\int\d r\,f(r)\ln\frac{f(r)}{\chi(r)}\;.
\label{eq:24}
\end{equation}
Here, $\chi(r)$ is a smooth prior function chosen to suppress small scale fluctuations. A suitable prior can be the smoothed version of the deprojection result from the previous iteration step. This choice, known as the floating default (see \citep{horne_maximum_1985, lucy_optimum_1994}), is defined by
\begin{equation}
  \chi(r) = \int\d r'\,P(r|r')f(r')\;,
\end{equation}
with a normalised, usually sharply peaked convolution kernel $P(r|r')$ symmetric in $r-r'$. We use a properly normalised Gaussian with a smoothing scale $L$,
\begin{equation} 
  P(r|r') \propto \exp\left(-\frac{(r-r')^2}{L^2}\right)\;.
\end{equation}

Replacing the variation of $H[f]$ in Eq.~(\ref{var1}) by the variation of $Q[f]$ from Eq.~(\ref{eq:regularised_functional}), we obtain
\begin{align}
\label{eq:iteration}
  f_{i+1}(r) &= f_i(r)\left\{
    \int\d s\,\frac{g(s)}{\tilde g_i(s)}K(s|r)\right.\\&+\left.
    \alpha\left[1+S[f_i]+\ln\left(\frac{f_i(r)}{\chi_{i}(r)}\right)-
    \int\d r'\,\frac{f_i(r')}{\chi_{i}(r')}P(r|r')\right]\right\}\;,\nonumber
\end{align}
which we shall use henceforth. Since we are working with discretised data sets, the integrals in Eq.~(\ref{eq:iteration}) need to be approximated by sums.
 
\section{Numerical tests}
\label{sect:num_tests}

We now proceed to demonstrate that it is possible to recover the projected gravitational potential $\psi$ of a galaxy cluster from the measured velocity dispersions projected along the line-of-sight. The algorithm described in Subsect.~\ref{subs:relations} assumes spherical symmetry and a polytropic relation between the effective galaxy pressure and the density.

Feeding the projected velocity dispersions into the Richardson-Lucy deprojection, we obtain the effective pressure $P$ related to the gravitational potential $\phi$ by Eq.~(\ref{eq:Volterra}). Once this is achieved, the gravitational potential just needs to be projected along the line-of-sight.

\subsection{The data}
\label{subs:data}

For testing our algorithm with simulated data, we require a velocity-dispersion profile projected along the line-of-sight and a two-dimensional gravitational potential obtained independently. We obtain such data from one of the clusters (denoted \textit{g1}) in the hydrodynamically simulated sample described in \citep{saro_cluster_g1_2006} and used previously in \citep{meneghetti_weighing_2010}. We start from a catalogue listing the Cartesian coordinates and the three Cartesian velocity components of simulation particles tracking the motion of cluster galaxies. All information necessary for the kinematic analysis described in Subsect.~\ref{subs:testing} can be extracted from this catalogue.

\begin{figure*}
  \subfigure[]{\includegraphics[width=0.48\hsize]{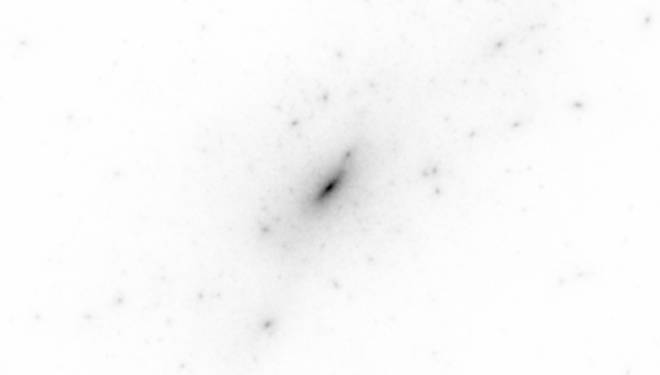}}\hfill
  \subfigure[]{\includegraphics[width=0.48\linewidth]{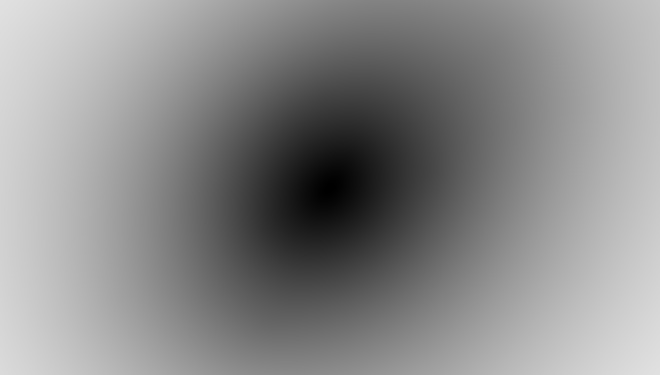}}
\caption{(a) Surface-mass density of the simulated cluster \textit{g1} in the $x$-$y$ plane. (b) Two-dimensional gravitational potential after obtained from the surface-density map solving Poisson's equation via fast-Fourier transform.}
\label{fig:maps}
\end{figure*}

In addition, we converted the surface-mass density of the cluster into the two-dimensional potential, solving Poisson's equation by means of a fast-Fourier transform. The surface-density map and the two-dimensional gravitational potential are shown in Fig.~\ref{fig:maps}. Evidently, the potential is considerably smoother than the mass density.

To enforce axial symmetry, we azimuthally average the gravitational potential shown in Fig.~\ref{fig:maps} (b) around the centre, chosen to be the point with the deepest (most negative) potential. In order to make it comparable to the outcome of our algorithm, it can be passed into the RL deprojection (with $\beta = 0$) yielding the gravitational potential which can then be appropriately shifted and normalised. 

\subsection{Testing the algorithm}\label{subs:testing}

Assuming spherical symmetry, we transform from Cartesian to spherical polar coordinates using
\begin{align}
 r &= \sqrt{x^2+y^2+z^2}\;,\nonumber\\
 v_r &= v_x\frac{x}{r} + v_y\frac{y}{r}+v_z\frac{z}{r}\;,\nonumber\\
 v_\phi &= -v_x\frac{y}{\sqrt{x^2+y^2}}+v_y\frac{x}{\sqrt{x^2+y^2}}\;,\nonumber\\
 v_\theta &= v_x\frac{xz}{r\sqrt{x^2+y^2}}+v_y\frac{yz}{r\sqrt{x^2+y^2}}-v_z\frac{\sqrt{x^2+y^2}}{r}\;.
\end{align}
The number-density profile of cluster galaxies is obtained within radial bins $r_i$ centred on the cluster centre chosen to contain equal numbers of galaxies. The number counts are then converted to a number-density profile by weighting with the inverse volume of each radial shell,
\begin{equation}
  n(r) \propto \frac{\mathrm{counts}}{r_{i+1}^3 - r_i^3}\;.
\end{equation}

\begin{figure}[ht]
  \includegraphics[width=\hsize]{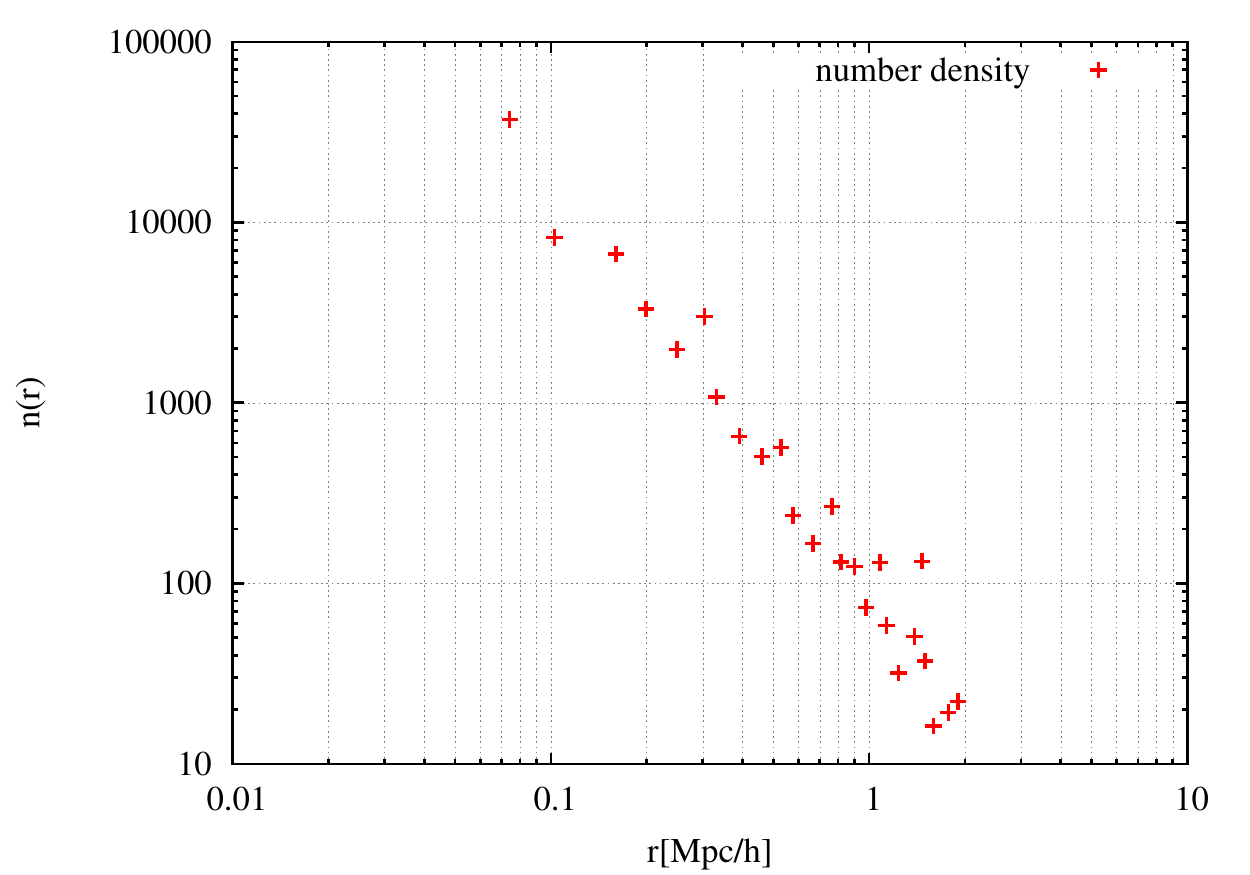}
\caption{The number density profile of the simulated cluster galaxies is show vs.\ the radius. It is obtained by galaxy number counts in spherical shells.}
\label{fig:number_density}
\end{figure}

Figure~\ref{fig:number_density} reveals that the number-density profile for this particular cluster essentially follows a power law. In a first approximation sufficient for our purposes, galaxy biasing and variations of the galaxy mass function are neglected, allowing us to adopt the number density as a direct tracer of the mass density.

\begin{figure}[ht]
  \includegraphics[width=\hsize]{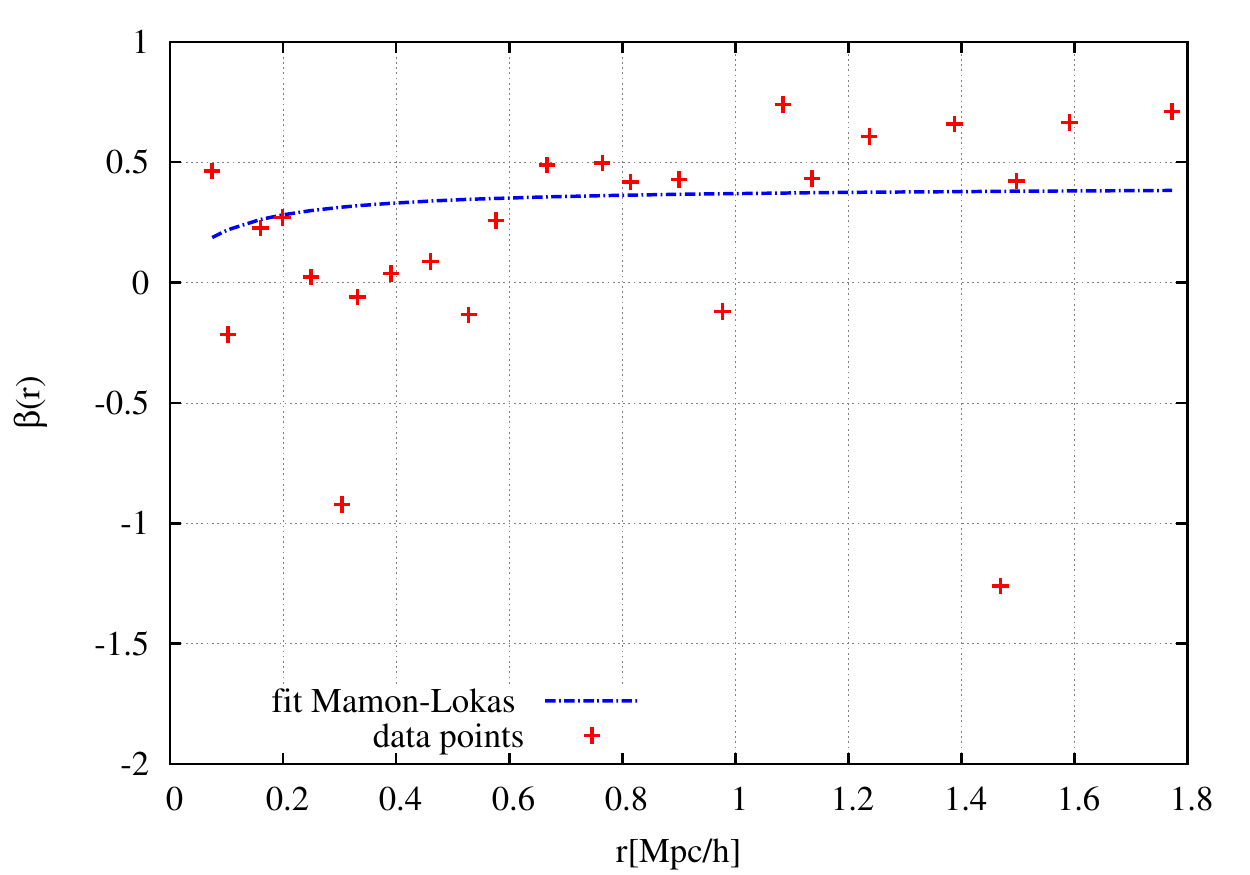}
\caption{Radial profile of the anisotropy parameter $\beta(r)$ defined in Eq.~(\ref{eq:beta_parameter}), obtained from the simulated cluster data. The two-parameter model by \citep{mamon_dark_2005} is fitted to the data points. We find $\beta_\infty = 0.40\pm0.5$ and $r_\beta = 0.08\pm0.74$. Concerning this result, a vanishing $\beta$ would also be allowed.}
\label{fig:beta_profile}
\end{figure}

Given the number-density profile and the radial bins, a mean galaxy velocity can be obtained in each radial shell. The variance of the velocities about this mean yields the galaxy velocity-dispersion profiles $\sigma_r^2$, $\sigma_\phi^2$ and $\sigma_\theta^2$ in the radial, azimuthal and polar directions. These quantities enable us to derive the profile of the anisotropy parameter according to the definition in Eq.~(\ref{eq:beta_parameter}). The result is shown in Fig.~\ref{fig:beta_profile}. Despite the substantial scatter, we fit the two-parameter model
\begin{equation}
  \beta(r) = \beta_\infty\frac{r}{r+r_\beta}
\end{equation}
proposed by \citep{mamon_dark_2005}, generalized by \citep{tiret_velocity_2007} and used in \citep{mamon_mamposst:_2013}. This fitted profile will be used as the mean anisotropy-parameter profile in the reconstruction, despite its being a rough approximation.

We weigh the radial velocity dispersions $\sigma_r^2$ with the number-density and obtain the effective galaxy pressure. Projecting this quantity along the line-of-sight,
\begin{equation}
  p_z = \int_s^{r_\mathrm{max}}\d r\,p_r(r)K(s|r)\;,
\label{los_p}
\end{equation}
with the projection kernel $K(s|r)$ defined in Eq.~(\ref{eq:kernel}), we finally complete the input for our method which would be a proper observable provided by observations. Note that the observed velocity dispersions are quantities projected along the line-of-sight and are thus implicitly weighted by the number density of galaxies along the line-of-sight.

\begin{figure}[ht]
  \includegraphics[width=\hsize]{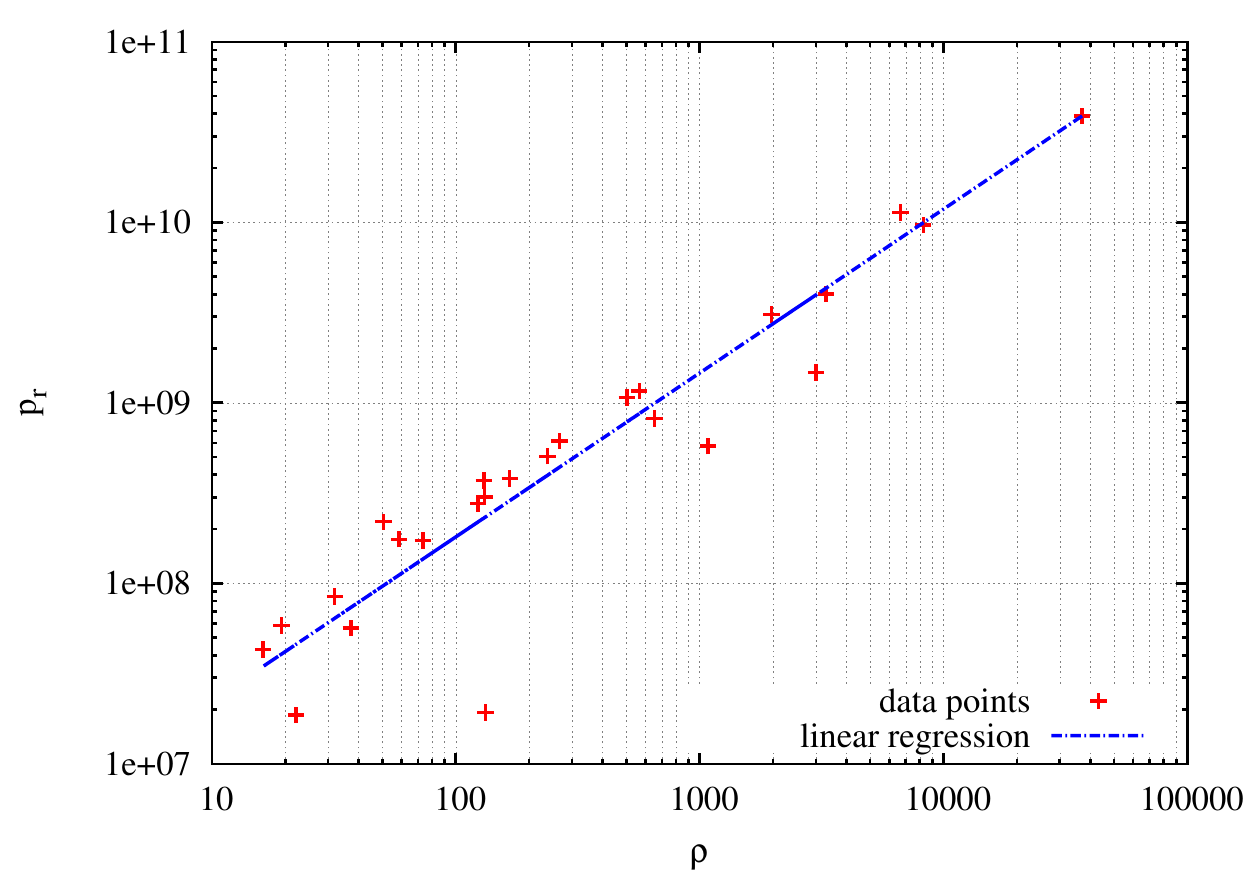}
\caption{The effective galaxy pressure is shown here vs.\ the density. The relation is well represented by a power law, supporting our assumption of an effective polytropic relation. The mean polytropic index, derived by linear regression, is $\gamma = 0.91\pm0.03$.}
\label{fig:pressure_vs_density}
\end{figure}

Figure~\ref{fig:pressure_vs_density} is a double-logarithmic plot of the effective galaxy pressure vs.\ the density, confirming that our polytropic assumption is reasonable. The mean polytropic index $\gamma$, introduced in Sect.~\ref{subs:relations}, is estimated from it by linear regression.

\begin{figure*}
  \subfigure[]{\includegraphics[width=0.48\linewidth]{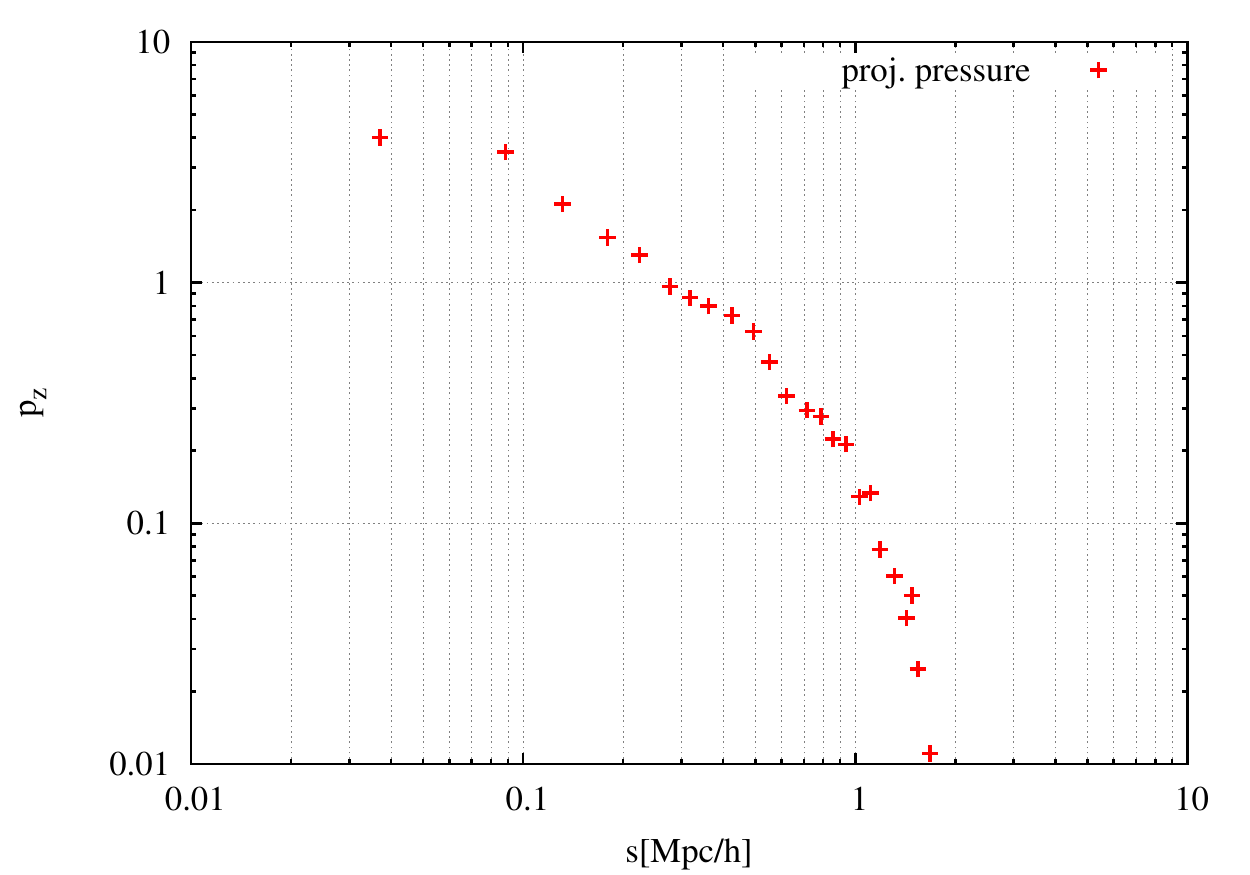}}\hfill
  \subfigure[]{\includegraphics[width=0.48\linewidth]{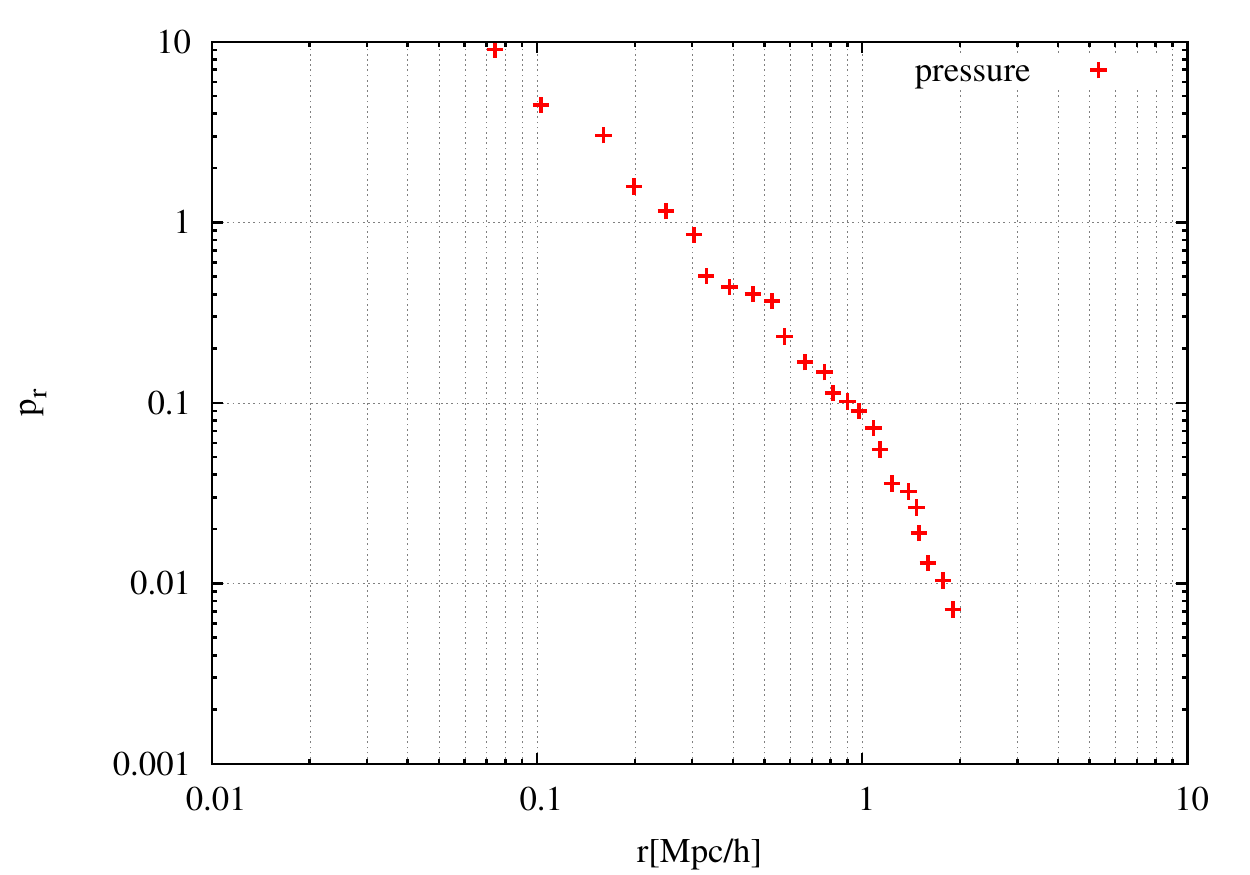}}\newline
  \subfigure[]{\includegraphics[width=0.48\linewidth]{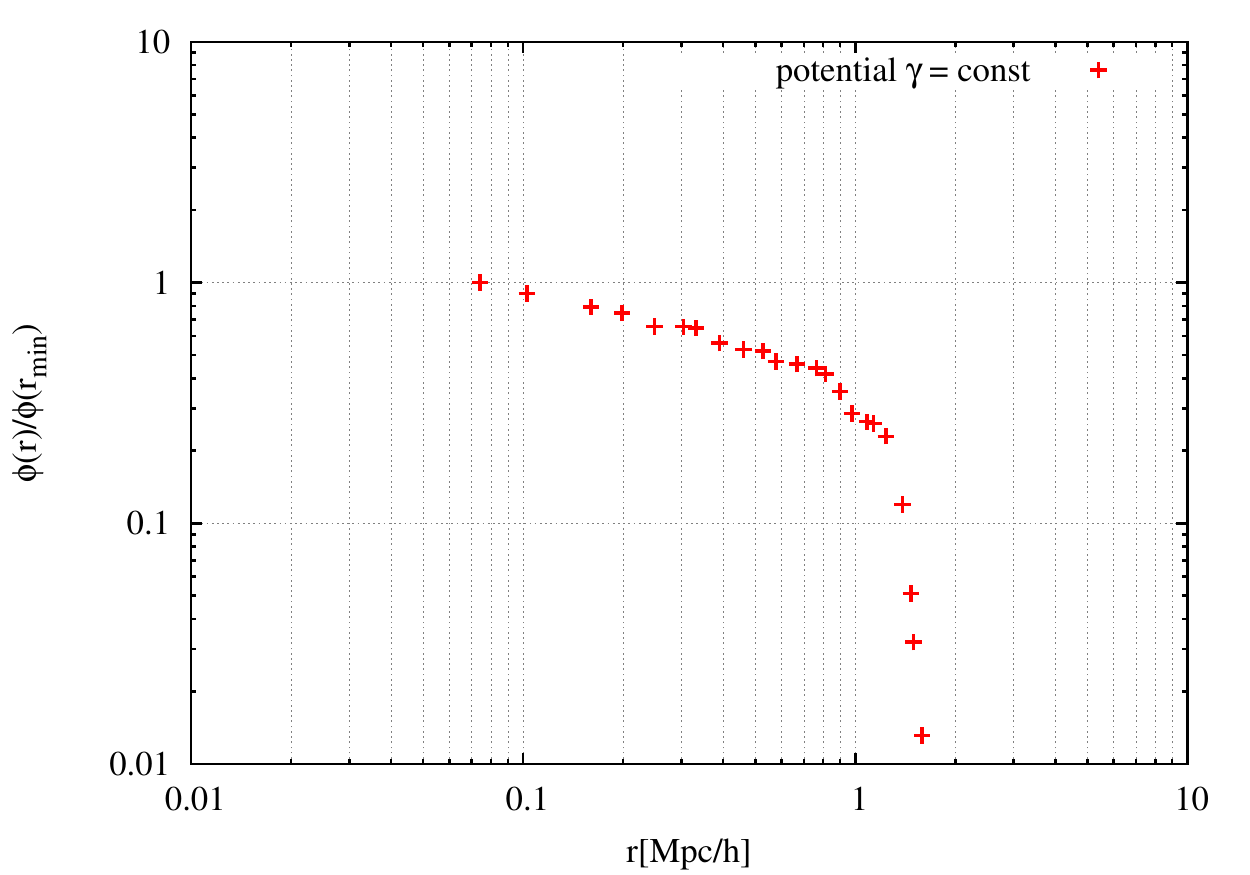}}\hfill
  \subfigure[]{\includegraphics[width=0.48\linewidth]{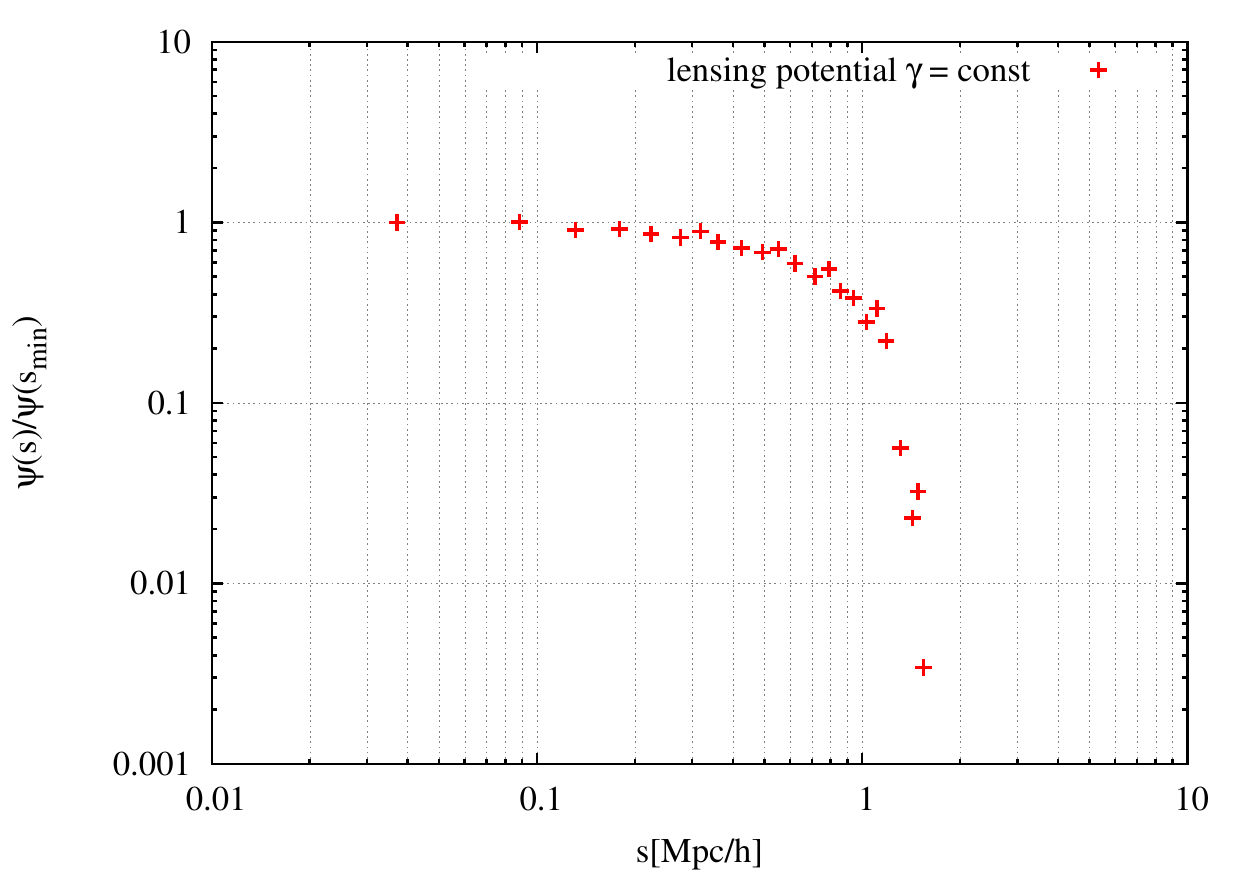}}
\caption{Results of the four different steps comprising our algorithm. (a) The input to our pipeline is the normalised, line-of-sight projected velocity-dispersion profile as a function of the projected radius and weighted by the galaxy number density. (b) The Richardson-Lucy deprojection algorithm yields an effective galaxy-pressure profile. (c) Solving the Volterra integral equation (Eq. (\ref{eq:Volterra})), we obtain the three-dimensional, Newtonian potential. (d) The last step consists in projecting the gravitational potential of panel (c) to find the two-dimensional potential.}
\label{fig:pipeline}
\end{figure*}

Figure~\ref{fig:pipeline} illustrates our complete algorithm. In the top-left panel, we show the normalised, line-of-sight projected velocity dispersions as a function of the projected radius and weighted with the galaxy number density. In the top-right panel, the normalised effective pressure profile obtained from the Richardson-Lucy deprojection algorithm is displayed. We then invert Eq.~(\ref{eq:Volterra}) to obtain the three-dimensional, Newtonian gravitational potential shown in the bottom-left panel. In the bottom-right panel, we finally show the derived two-dimensional gravitational potential.

Figure~\ref{fig:pipeline} provides an overview over the first test we performed on our algorithm and shows all steps needed to reconstruct the two-dimensional potential from the line-of-sight projected velocity dispersions. A comparison between our reconstructed profiles and the true three- and two-dimensional gravitational potentials is now in order. We define as ``true'' the potentials extracted directly from the numerically-simulated cluster.

\begin{figure*}[!ht]
  \subfigure[]{\includegraphics[width=0.48\linewidth]{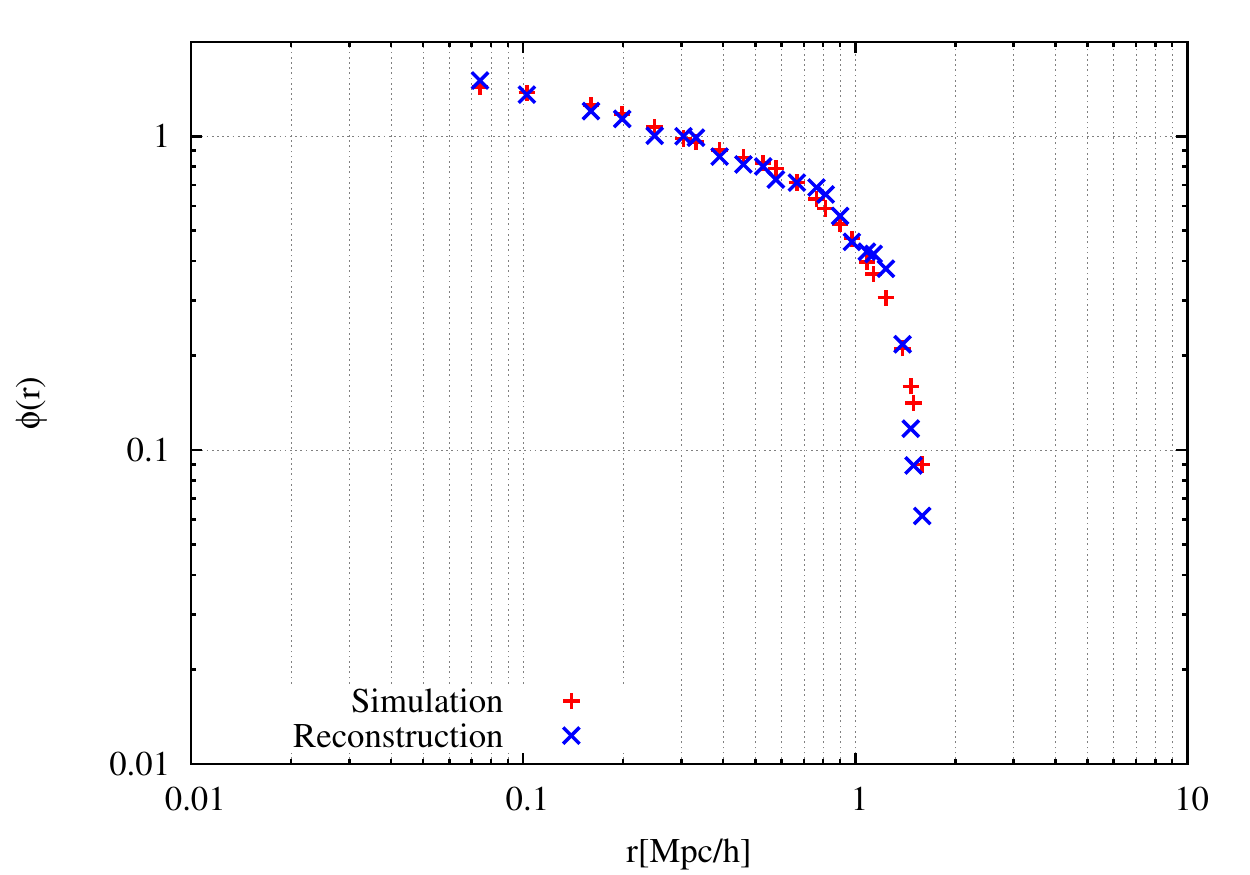}}\hfill \subfigure[]{\includegraphics[width=0.48\linewidth]{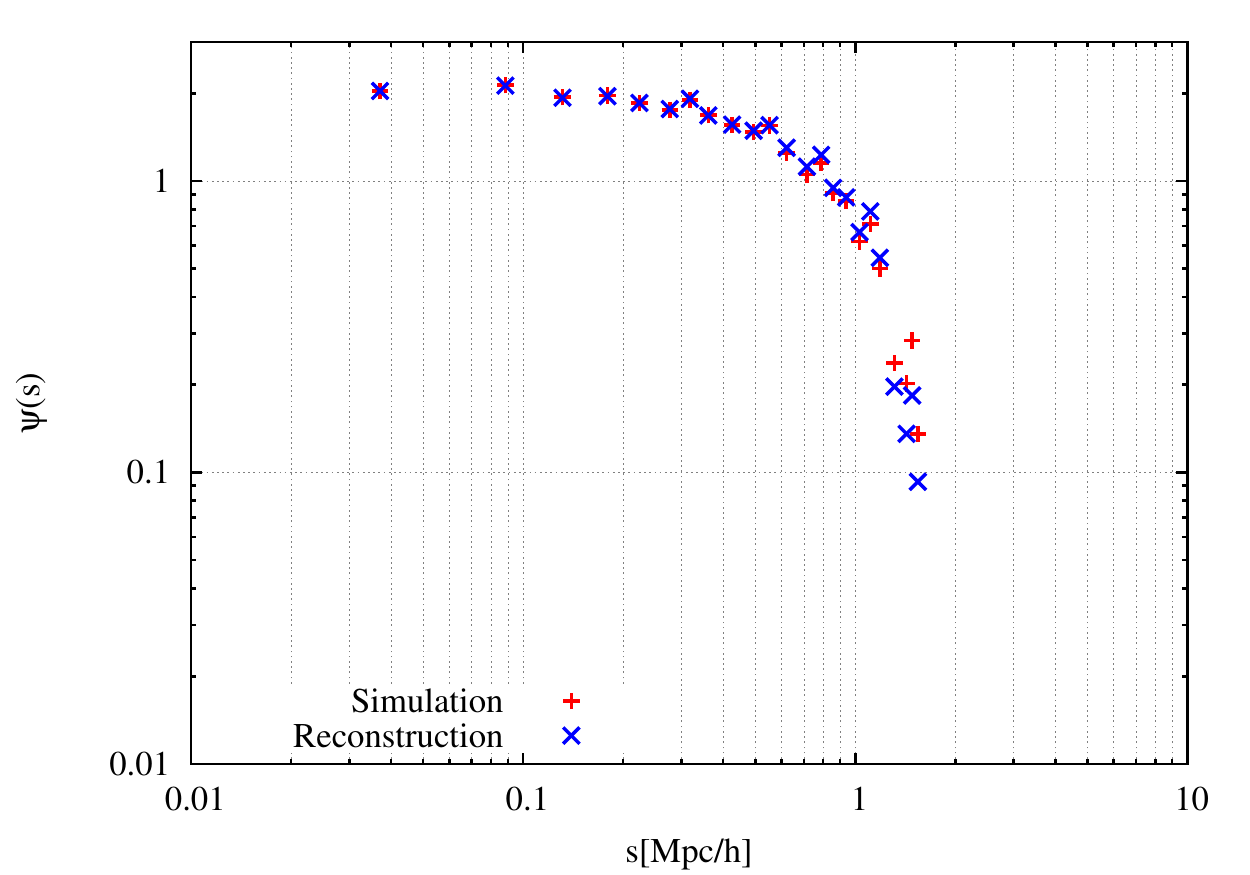}}
\caption{The reconstructed gravitational potentials in (a) three and (b) two dimensions are plotted as functions of radius and compared with the true potentials. In each panel, the blue points show the true potential determined from the convergence map, while the red points show the result of our reconstruction method. The corresponding relative deviation as a function of radius is shown in Fig.~\ref{mean squared deviation}.}
\label{fig:comparison}
\end{figure*}

\begin{figure}
  \includegraphics[width=\hsize]{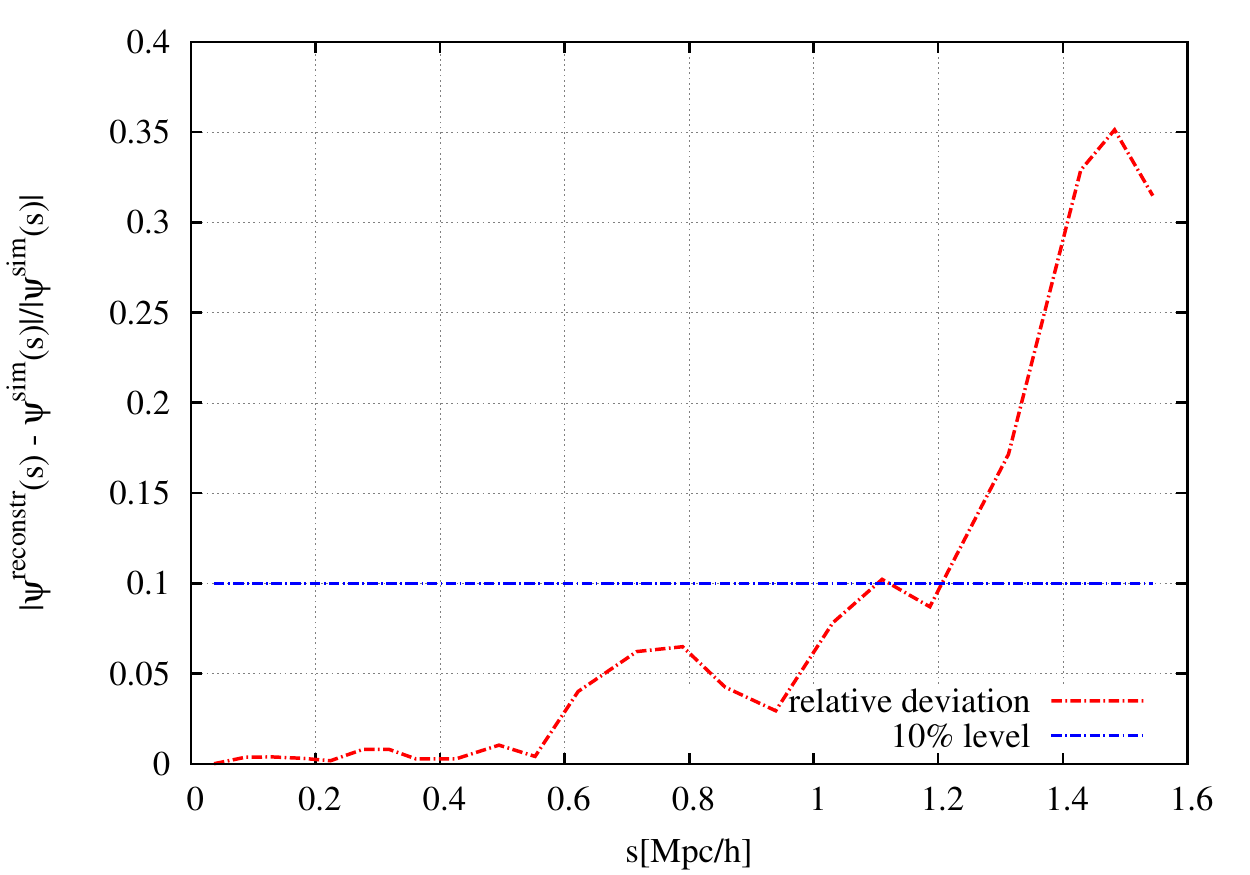}
\caption{The relative deviation between the reconstructed and true two-dimensional gravitational potentials is shown here as a function of distance from the cluster center. The deviation remains moderate (below $10\,\%$) within a radius of approximately $1.2\,h^{-1}\mathrm{Mpc}$.} 
\label{mean squared deviation}
\end{figure}

Figure~\ref{fig:comparison} (a) shows the comparison between the reconstructed and the true gravitational potentials, while Fig.~\ref{fig:comparison} (b) compares the reconstructed and the true projected potentials. In both panels, the blue curve shows the true potential extracted directly from the simulation, while the red curve shows our reconstruction results. For easier comparison of both potentials, their zero points and normalisations are adjusted\footnote{Note that due to the normalisation constraint of the Richardson-Lucy deprojection, we obtain a function proportional to $\phi$. The potential extracted from the simulation has to be normalised in the same way. Since both potentials are obtained from different datasets providing different spatial boundaries, they need to be appropriately shifted (gauged) by adding a constant factor.}. To be consistent with \citep{konrad_X_rays_2013} and \citep{majer_SZ_2013}, we decided to normalise both functions to unity and fix $\phi(r_\mathrm{max}) = 0$. 
In Fig.~\ref{mean squared deviation}, the relative deviation
\begin{equation}
  \Delta\psi(s) = \frac{\left|\psi^\mathrm{reconstr}(s) - \psi^\mathrm{sim}(s)\right|}
  {\left|\psi^\mathrm{sim}(s)\right|}\;,
\end{equation}
 between the reconstructed and the true two-dimensional potentials is shown as a function of projected radius. The deviation strongly increases at large radii, although it remains below $10\,\%$ within $1.2\,h^{-1}\mathrm{Mpc}$, which is similar to the virial radius of the cluster.

\subsection{Application to observed data}

The numerical tests presented above were run on line-of-sight projected data extracted from a simulated galaxy cluster. Starting from spatial Cartesian coordinates and velocity components of a sample of member galaxies, the effective galaxy pressure was obtained by weighting the squared radial velocity dispersion with the number density. Afterwards, the required observable was constructed by projection along the line-of-sight. The effective polytropic index $\gamma$ and the profile $\beta(r)$ of the anisotropy parameter could be easily obtained from the simulated sample. Unfortunately, none of the latter two quantities is directly accessible to observations. Therefore a method needs to be proposed that allows to model the effective polytropic index and the anisotropy profile.

Major efforts have been undertaken in recent years to either constrain the anisotropy profile from observational and simulated data \citep{lemze_dynamical_2009, mamon_mamposst:_2013} or to identify a general relation describing it \citep{mamon_dark_2005, hansen_universal_2006}. The main objective of our work is the reconstruction of the projected gravitational potential of a galaxy cluster. However, we can envisage an inverse application of our algorithm. By an operation analogous to \citep{lemze_dynamical_2009}, it is possible to constrain $\beta(r)$ using simultaneously information from gravitational lensing and galaxy kinematics. The existence of multiple constraints on the two-dimensional potential $\psi$ also allows developing an iterative method for constraining $\beta(r)$ and $\gamma$.

Within the CLASH survey project, considerable progress has been made in reproducing the observed CLASH clusters by simulations (Meneghetti et al. 2013, in prep.). These results would possibly enable us to guess a polytropic index and an anisotropy profile which could then be combined with kinematic data from observations in order to improve our reconstruction.

\section{Conclusions}\label{sect:Conclusions}

With the study presented here, we have continued our series of papers aiming at parameter-free reconstructions of projected gravitational potential of galaxy clusters. The essential goal of our approach is thus to construct methods to relate all cluster observables to the two-dimensional gravitational potential. In two preceding papers, we have shown how this can be achieved with the two cluster observables based on the hot intracluster gas, i.e.\ X-ray emission and the thermal Sunyaev-Zel'dovich effect. In this paper, we address the problem how the observed line-of-sight projected velocity dispersions of cluster galaxies can be converted to the projected gravitational potential. We propose a solution which follows closely our interpretation of the observables provided by the intracluster plasma, but necessarily differs from it in the crucial aspect that the velocity dispersion of the cluster galaxies, unlike the gas pressure, can be anisotropic.

The algorithm we propose rests on the following crucial assumptions. First, we assume that the effective galaxy pressure, by which we mean the product of the galaxy number density and the radial velocity dispersion, can be related to the density itself by a polytropic relation, i.e.\ a power law. We test this assumption with different density profiles and functional forms of the anisotropy parameter and find it well satisfied. Second, adopting this polytropic relation, we solve the radial component of the Jeans equation, relating the effective galaxy pressure to the potential gradient. This solution can be analytically given in the form of a Volterra integral equation of the first kind for the three-dimensional gravitational potential, which can be solved by iteration. Thus, we establish a relation between the three-dimensional gravitational potential and the effective galaxy pressure.

The effective galaxy pressure itself can be obtained from the observable, line-of-sight projected galaxy velocity dispersions by Richardson-Lucy deprojection. The two-dimensional gravitational potential can finally be found by straightforward projection along the line-of-sight.

We have tested this algorithm on a simulated galaxy cluster in which galaxies have been identified. This cluster is part of a sample of numerical hydrodynamical simulations described in \citep{saro_cluster_g1_2006} and used in \citep{meneghetti_weighing_2010}. The anisotropy profile $\beta(r)$, as well as the mean effective polytropic index $\gamma$, were obtained directly from the simulation. Although these quantities are hardly directly measurable from observations and it was up to now impossible to give general prescriptions of their behaviour, it may well be justified to calibrate them on numerical simulations without introducing an unacceptable bias.

The numerical simulation shows that the anisotropy profile shown in Fig.~\ref{fig:beta_profile} may be poorly constrained by the given simulation. The fit performed by the model suggested by \citep{mamon_dark_2005} covers a wide range of possible behaviour. Nonetheless, reconstructing the two-dimensional potential based on the nominal fit yields a result that closely follows the numerical expectation out to $\approx1.2\,h^{-1}\mathrm{Mpc}$, as demonstrated by the relative deviation of the reconstructed from the true projected gravitational potential (see Fig.~\ref{mean squared deviation}). The deviation remains below $10~\%$ in this radial range. Figure~\ref{fig:pressure_vs_density} confirms the approximate treatment of our system as a polytropic stratification.

The main limitation of our algorithm so far lies in the assumption of spherical symmetry, which we plan to relax by extending our method to the axially (rather than spherically) symmetric case.

We are planning to proceed by applying the algorithm devised here to observational data that could possibly be provided by the CLASH survey. In addition, we will focus on unifying the three algorithms presented in \citep{konrad_X_rays_2013, majer_SZ_2013} and this paper such as to allow a single, joint reconstruction of the two-dimensional potential of galaxy clusters using lensing, X-ray, Sunyaev-Zel'dovich and kinematic data.

\acknowledgements{This work was supported in part by contract research `Internationale Spitzenforschung II-1' of the Baden-W\"urttemberg Stiftung, by the Collaborative Research Centre TR~33 and project BA 1369/17 of the Deutsche Forschungsgemeinschaft. ES wishes to warmly thank Lauro Moscardini for his kind hospitality in Bologna and for the useful and insightful discussions.}

\bibliographystyle{aa}
\bibliography{references_corr}

\end{document}